\documentstyle[epsf,eqsecnum,floats,preprint,aps]{revtex}

\tighten

\input epsf
\begin{document}

\newcommand{\be}{\begin{equation}}
\newcommand{\ee}{\end{equation}}
\newcommand{\bea}{\begin{eqnarray}}
\newcommand{\eea}{\end{eqnarray}}
\newcommand{\PSbox}[3]{\mbox{\rule{0in}{#3}\includegraphics{#1}\hspace{#2}}}

\def\5M{M^3_{(5)}}
\def\4M{M^2_{(4)}}

\overfullrule=0pt
\def\Int{\int_{r_H}^\infty}
\def\d{\partial}
\def\e{\epsilon}
\def\M{{\cal M}}
\def\high{\vphantom{\Biggl(}\displaystyle}
\catcode`@=11
\def\@versim#1#2{\lower.7\p@\vbox{\baselineskip\z@skip\lineskip-.5\p@
    \ialign{$\m@th#1\hfil##\hfil$\crcr#2\crcr\sim\crcr}}}
\def\simge{\mathrel{\mathpalette\@versim>}} %
\def\simle{\mathrel{\mathpalette\@versim<}} %
\def\sun{\hbox{$\odot$}}
\catcode`@=12 

\rightline{CWRU--P18--02}
\rightline{astro-ph/0212083}
\vskip 4cm

\setcounter{footnote}{0}

\begin{center}
\large{\bf Gravitational Leakage into Extra Dimensions:\\
Probing Dark Energy Using Local Gravity}
\ \\
\ \\
\normalsize{Arthur Lue\footnote{E-mail:  lue@bifur.cwru.edu}
  and Glenn Starkman\footnote{E-mail:  starkman@balin.cwru.edu}}
\ \\
\ \\
\small{\em Department of Physics\\
Case Western Reserve University \\
Cleveland, OH 44106--7079}

\end{center}

\begin{abstract}

\noindent
The braneworld model of Dvali--Gabadadze--Porrati (DGP) is a theory
where gravity is modified at large distances by the arrested leakage
of gravitons off our four-dimensional universe.  Cosmology in this
model has been shown to support both ``conventional" and exotic
explanations of the dark energy responsible for today's cosmic
acceleration.  We present new results for the gravitational field of a
clustered matter source on the background of an accelerating universe
in DGP braneworld gravity, and articulate how these results differ
from those of general relativity.  In particular, we show that orbits
nearby a mass source suffer a universal anomalous precession as large
as $\pm 5~\mu{\rm as/year}$, dependent only on the graviton's
effective linewidth and the global geometry of the full,
five-dimensional universe.  Thus, this theory offers a local gravity
correction sensitive to factors that dictate cosmological history.
\end{abstract}

\setcounter{page}{0}
\thispagestyle{empty}
\maketitle

\eject

\vfill

\baselineskip 18pt plus 2pt minus 2pt

\section{Introduction}

The gravity theory of Dvali--Gabadadze--Porrati (DGP) is a braneworld
theory with a metastable four-dimensional graviton
\cite{Dvali:2000hr}.  The graviton is pinned to a four-dimensional
braneworld by intrinsic curvature terms induced by quantum matter
fluctuations; but as it propagates over large distances, the graviton
eventually evaporates off the brane into an infinite volume,
five-dimensional Minkowski bulk.  There exists a single free parameter
in DGP braneworld gravity, the crossover scale, $r_0$.  This scale
dictates that distance below which gravity is controlled by brane
effects, but larger than which gravity assumes a five-dimensional
behavior.  As a result, the DGP braneworld theory is a model in a
class of theories in which gravity deviates from conventional Einstein
gravity not at short distances (as in more familiar braneworld
theories), but rather at long distances.  Such a model has both
intriguing phenomenological \cite{Dvali:2001gm,Dvali:2001gx} as well
as cosmological consequences
\cite{Deffayet,Deffayet:2001pu,Deffayet:2002sp,Alcaniz:2002qh,Jain:2002di,Alcaniz:2002qm,Lue:2002fe}.
A braneworld model of the sort where gravity is modified at extremely
large scales is motivated by the desire to ascertain how our
understanding of cosmology may be refined by the presence of extra
dimensions.  Indeed, there exist novel cosmologies in this theory that
provide an alternative explanation of the cosmic dark energy
\cite{Deffayet}.

Recovery of Einstein gravity at short distance scales in DGP
\cite{Deffayet:2001uk}, especially for static sources
\cite{Lue:2001gc,Gruzinov:2001hp,Porrati:2002cp,Middleton:2002qa}, is
a subtle effect.  Even though gravity is four-dimensional at distances
less than $r_0$, it is not always Einstein gravity.  
For a point source whose Schwarzschild radius is $r_g$,  
general relativity is only recovered for distances shorter than
\be
	r_* = \left(r_0^2r_g\right)^{1/3}\ .
\ee 
For distances larger than $r_*$, gravity is four-dimensional
linearized Brans--Dicke, with parameter $\omega = 0$.  Thus, a marked
departure from conventional physics continues down to distances much
smaller than $r_0$, the distance at which the extra dimension is
naively hidden.  That departure is well-characterized and provides a
possible signature for the existence of extra dimensions.

There is, however, a catch.  The cosmological solutions that drive
interest in DGP gravity indicate that $r_0$ should be close to today's
Hubble radius.  Localized matter sources are embedded in a
cosmological spacetime.  Far enough away from a given source, the
motion of observers and test particles is dominated by the cosmology,
rather than the gravity of the matter source.  However, if $r_0$ is
today's Hubble radius, then the distance away from a source at which
cosmology dominates the metric is also $r_*$.  So, the DGP departure
from Einstein gravity that was calculated in a static Minkowski
background is only significant on length scales where the background
cannot actually be approximated as Minkowski.  The calculations needs
to be refined if we are to have the correct new physics signature.

The subject of this paper is to look at corrections to Einstein
gravity for the DGP braneworld theory in a cosmological background.\footnote{
  The work in this paper, as well as that in
  Refs.~\cite{Lue:2001gc,Gruzinov:2001hp,Porrati:2002cp}, addresses
  issues similar to those studied in
  Refs.~\cite{Kofinas:2001qd,Kofinas:2002gq}.  Our results differ from
  those in the latter work because we are only concerned with solutions
  that asymptotically approach well-behaved, familiar solutions (e.g.,
  Minkowski space or deSitter expansion) far away from the matter source,
  including far away from the source in the
  direction into the bulk.}
We begin by laying out the necessary details of the model as well as
the particular background of interest.  We then solve for the metric
of a spherically symmetric, static matter source in that background.
We indeed find that the corrections to Einstein are sensitive to the
background cosmology, even in the region inside $r_*$ where the
cosmological flow is ostensibly irrelevant.  We discuss possible tests
for the detection of new physics at astronomical scales and suggest
that it is possible for such tests of local gravity to reveal
information about global features of the full five-dimensional
cosmology, and ultimately shed light on the nature of dark energy and
today's cosmic expansion.

\section{Preliminaries}

\subsection{The Model}

Consider a braneworld theory of gravity with an infinite-volume bulk
and a metastable brane graviton \cite{Dvali:2000hr}.  We take a
four-dimensional braneworld embedded in a five-dimensional Minkowski
spacetime.  The bulk is empty; all energy-momentum is isolated on the
brane.  The action is
\be
S_{(5)} = -\frac{1}{16\pi}M^3 \int d^5x
\sqrt{-g}~R +\int d^4x \sqrt{-g^{(4)}}~{\cal L}_m + S_{GH}\ .
\label{action}
\ee
The quantity $M$ is the fundamental five-dimensional Planck scale.  The
first term in Eq.~(\ref{action}) corresponds to the Einstein-Hilbert
action in five dimensions for a five-dimensional metric $g_{AB}$ (bulk
metric) with Ricci scalar $R$.  The term $S_{GH}$ is the Gibbons--Hawking
action.  In addition, we consider an intrinsic
curvature term which is generally induced by radiative corrections by
the matter density on the brane \cite{Dvali:2000hr}:
\be
-\frac{1}{16\pi}M^2_P \int d^4x \sqrt{-g^{(4)}}\ R^{(4)}\ .
\label{action2}
\ee
Here, $M_P$ is the observed four-dimensional Planck scale (see
\cite{Dvali:2000hr,Dvali:2001gm,Dvali:2001gx} for details).
Similarly, Eq.~(\ref{action2}) is the Einstein-Hilbert action for the
induced metric $g^{(4)}_{\mu\nu}$ on the brane, $R^{(4)}$ being its
scalar curvature.  The induced metric is\footnote{
	Throughout this paper, we use $A,B,\dots = \{0,1,2,3,5\}$ as
	bulk indices, $\mu,\nu,\dots = \{0,1,2,3\}$ as brane spacetime
	indices, and $i,j,\dots = \{1,2,3\}$ as brane spatial indices.}
\be
g^{(4)}_{\mu\nu} = \partial_\mu X^A \partial_\nu X^B g_{AB}\ ,
\label{induced}
\ee
where $X^A(x^\mu)$ represents the coordinates of an event on the brane
labeled by $x^\mu$.  The action given by Eqs.~(\ref{action}) and
(\ref{action2}) leads to the following equations of motion
\be
{1\over 2r_0}G_{AB} + \delta({\rm brane})G_{AB}^{(4)}
= {8\pi\over M_P^2}T_{AB}|_{\rm brane}\ ,
\label{Einstein}
\ee
where $G_{AB}$ is the bulk Einstein tensor, $G_{AB}^{(4)}$ is the
intrinsic brane Einstein tensor, and $T_{AB}|_{\rm brane}$ is the
matter energy-momentum tensor on the brane, and we have defined a
crossover scale
\be
	r_0 = {M_P^2 \over 2M^3}\ .
\label{r0}
\ee
This scale characterizes that distance over which metric fluctuations
propagating on the brane dissipate into the bulk \cite{Dvali:2000hr}.

\subsection{Field Equations}

We are interested in finding the metric for static, compact, spherical
sources.  We are interested in looking at these solutions in a
cosmological background rather than a Minkowski background
\cite{Gruzinov:2001hp,Porrati:2002cp} to ascertain what affects
cosmology might have on the observability of corrections to Einstein
gravity.  We restrict ourselves to a background deSitter cosmology.
Not only is such a background the simplest and most convenient, but
observations suggest that we are currently undergoing deSitter-like
cosmic acceleration.  If the Hubble scale of such an acceleration is
varying slowly, the results obtained here would apply.  They will also
shed light on the technical VDVZ problem and how one recovers Einstein
gravity in a deSitter background.

Under this circumstance of a static spherical source in a deSitter
background, one can choose a coordinate system in which the
cosmological metric is static (i.e., has a timelike Killing vector)
while still respecting the spherical symmetry of the matter source.
Let the line element be
\be
ds^{2} = N^2(r,z) dt^{2}
         - A^2(r,z)dr^2 - B^2(r,z)[d\theta^2 + \sin^2\theta d\phi^2]-dz^{2}\ .
\label{metric}
\ee
This is the most general static metric with spherical symmetry on the
brane.  The bulk Einstein tensor for this metric is:
\ \\
\bea
G_t^t &=& {1\over B^2}
-{1\over A^2}\left[{2B''\over B} - 2{A'\over A}{B'\over B}
  + {B'^2\over B^2}\right]
-\left[{\ddot{A}\over A} + {2\ddot{B}\over B}
  + 2{\dot{A}\over A}{\dot{B}\over B} + {\dot{B}^2\over B^2}\right]
\nonumber  \\
G_r^r &=& {1\over B^2}
-{1\over A^2}\left[2{N'\over N}{B'\over B} + {B'^2\over B^2}\right]
- \left[{\ddot{N}\over N} + {2\ddot{B}\over B}
  + 2{\dot{N}\over N}{\dot{B}\over B} + {\dot{B}^2\over B^2}\right]
\nonumber  \\
G_\theta^\theta &=& G_\phi^\phi =
-{1\over A^2}\left[{N''\over N} + {B''\over B}
  - {N'\over N}{A'\over A}+{N'\over N}{B'\over B}-{A'\over A}{B'\over B}\right]
  \nonumber \\
&& \ \ \ \ \ \ \ \ \ \ \ \ \ \ \ \ \ \ \ \ \ \ \ \ \ \ \ \ \ \ \ \ \ \ \ \ \ \ 
\ \ \ \ \ \ \ 
- \left[{\ddot{N}\over N} + {\ddot{A}\over A} + {\ddot{B}\over B}
  + {\dot{N}\over N}{\dot{A}\over A} + {\dot{N}\over N}{\dot{B}\over B}
  + {\dot{A}\over A}{\dot{B}\over B}\right]
\label{5dEinstein}\\
G_z^z &=& {1\over B^2}
-{1\over A^2}\left[{N''\over N} + {2B''\over B}
  - {N'\over N}{A'\over A}+2{N'\over N}{B'\over B}-2{A'\over A}{B'\over B}
  + {B'^2\over B^2}\right]     \nonumber \\
&& \ \ \ \ \ \ \ \ \ \ \ \ \ \ \ \ \ \ \ \ \ \ \ \ \ \ \ \ \ \ \ \ \ \ \ \ \ \
 \ \ \ \ \ \ \ \ \ \ \ \ \ \ \ \ \ 
- \left[{\dot{N}\over N}{\dot{A}\over A} + 2{\dot{N}\over N}{\dot{B}\over B}
  + 2{\dot{A}\over A}{\dot{B}\over B} + {\dot{B}^2\over B^2}\right]
 \nonumber  \\
G_{zr} &=& -\left[{\dot{N}'\over N} + {2\dot{B}'\over B}\right] +
{\dot{A}\over A}\left({N'\over N} + {2B'\over B}\right)\ . \nonumber
\eea
\ \\
The prime denotes partial differentiation with respect to $r$,
whereas the dot represents partial differentiation with respect to $z$.

We wish to solve the five-dimensional field equations,
Eq.~(\ref{Einstein}).  This implies that all components of the
Einstein tensor, Eqs.~(\ref{5dEinstein}), vanish in the bulk but
satisfy the following modified boundary relationships on the brane.
Fixing the residual gauge $B|_{z=0}=r$, when $z=0$ and imposing
${\cal Z}_2$--symmetry across the brane
\bea
-\left({\dot{A}\over A} + {2\dot{B}\over B}\right)
&=& {r_0\over A^2}\left[-{2\over r}{A'\over A} + {1\over r^2}(1-A^2)\right]
+ {8\pi r_0\over M_P^2}\rho(r)
\nonumber \\
-\left({\dot{N}\over N} + {2\dot{B}\over B}\right)
&=&  {r_0\over A^2}\left[{2\over r}{N'\over N} + {1\over r^2}(1-A^2)\right]
- {8\pi r_0\over M_P^2}p(r)
\label{branebc}\\
-\left({\dot{N}\over N} + {\dot{A}\over A} + {\dot{B}\over B}\right)
&=&  {r_0\over A^2}\left[{N''\over N} - {N'\over N}{A'\over A}
+ {1\over r}\left({N'\over N} - {A'\over A}\right)\right]
- {8\pi r_0\over M_P^2}p(r)\ ,
\nonumber
\eea
from $G_{tt}$, $G_{rr}$, and $G_{\theta\theta}$, respectively.

\newpage

\subsection{Background Cosmology}

The deSitter solution with Hubble scale, $H$, has the following
metric components:\footnote{
  The upper and lower signs in Eqs.~(\ref{deS-N}--\ref{deS-B})
  correspond to the two distinct cosmological phases that may exist
  for this theory \cite{Deffayet}.  The upper sign corresponds to the
  choice of Friedmann--Lema\^{\i}tre--Robertson--Walker (FLRW)
  cosmological phase, where a bulk observer views the braneworld as a
  relativistically expanding bubble from the interior; the lower sign
  corresponds to the self-accelerating cosmological phase, where the
  bulk observer views the braneworld as a relativistically expanding
  bubble from the exterior \cite{Deffayet,Lue:2002fe}.  These phases
  are vastly different geometrically and may have drastically
  different cosmological histories.  We use this sign convention
  throughout the paper.}
  
\bea
N(r,z) &=& (1 \mp H|z|)\left(1-H^2r^2\right)^{1/2}
\label{deS-N}     \\
A(r,z) &=& (1 \mp H|z|)\left(1-H^2r^2\right)^{-1/2}
\label{deS-A}    \\
B(r,z) &=& (1 \mp H|z|)~r\ .
\label{deS-B}
\eea
The brane energy-momentum tensor required for
a given Hubble parameter is
\be
T^A_B|_{\rm brane}= ~\delta (z)\ {\rm diag}
\left(\rho_H,-p_H,-p_H,-p_H,~0 \right)\ ,
\label{deS-EM}
\ee
with
\be
	\rho_H = -p_H = {3M_P^2H\over 8\pi r_0}\left(r_0H \pm 1\right)\ .
\label{deS-rho}
\ee
For physical reasons, we restrict ourselves to both non-negative
$\rho_H$ as well as non-negative $H$.  Under these circumstances,
one can see that $H \ge r_0^{-1}$ in the self-accelerating phase.
One can show that the solution Eqs.~(\ref{deS-N}--\ref{deS-rho})
corresponds to a coordinate transformation of the deSitter solution in
homogeneous cosmological coordinates found in Ref.~\cite{Deffayet}.

\section{Spacetime Geometry}

We have chosen a coordinate system, Eq.~(\ref{metric}), in which a
compact spherical matter source may have a static metric, yet still
exist within a background cosmology that is nontrivial (i.e., deSitter
expansion).  Let us treat the matter distribution to be that required
for the background cosmology, Eq.~(\ref{deS-EM}--\ref{deS-rho}), and
add to that a compact spherically symmetric matter source, located on
the brane around the origin ($r=0,z=0$)
\be
T^A_B|_{\rm brane}= ~\delta (z)\ {\rm diag}
\left(\rho_g(r)+\rho_H,-p_g(r)+\rho_H,
-p_g(r)+\rho_H,-p_g(r)+\rho_H,~0 \right)\ ,
\label{matter-EM}
\ee
where $\rho_g(r)$ is some given function of interest and $p_g(r)$
is chosen to ensure the matter distribution and metric are static.

We are interested only in weak matter sources, $\rho_g(r)$.
Moreover, we are most interested in those parts of spacetime where
deviations of the metric from Minkowski are small.  Then, it is
convenient to define the functions $\{n(r,z),a(r,z),b(r,z)\}$ such
that
\bea
N(r,z) &=& 1+n(r,z)     \\
A(r,z) &=& 1+a(r,z)
\label{linearize}      \\
B(r,z) &=& r~[1+b(r,z)]\ .
\eea
We may rewrite Eqs.~(\ref{5dEinstein}) and (\ref{branebc}) using the
functions $\{n(r,z),a(r,z),b(r,z)\}$.

In order to determine the metric on the brane, we will implement the
approximation
\be
\dot{n}|_{z=0} = \mp H\ ,
\label{assumption}
\ee
even in the presence of a compact matter source.
Equation~(\ref{assumption}) presumes that the contribution to
$\dot{n}|_{z=0}$ from the matter source is negligible compared to
other terms in the brane boundary conditions, Eqs.~(\ref{branebc}).
With this one specification, a complete set of equations,
represented by the brane boundary conditions Eqs.~(\ref{branebc}) and
$G_z^z=0$, exists on the brane so that the metric functions may be
solved on that surface without reference to the bulk.  We justify this
assumption by examining the full bulk metric in
detail in the Appendix.

The brane boundary conditions Eqs.~(\ref{branebc}) now take the form
\bea
-(\dot{a}+2\dot{b}) &=& r_0\left[-{2a' \over r} - {2a\over r^2}\right]
			+ {r_0\over r^2}R'_g(r) + 3H(r_0H \pm 1)
\nonumber    \\
-2\dot{b} &=& r_0\left[{2n' \over r} - {2a\over r^2}\right]
	                + H(3r_0H \pm 2)
\label{branebc2}     \\
-(\dot{a} + \dot{b}) &=& r_0\left[n'' + {n'\over r} - {a'\over r} \right]
	                + H(3r_0H \pm 2)\ ,
\nonumber
\eea
where we have defined an effective Schwarzschild radius $R_g(r)$ for the
matter source inside a given distance, $r$, 
\be
R_g = {8\pi\over M_P^2}\int_0^r dr~r^2\rho_g(r)\ ,
\ee
and where we have neglected second-order contributions (including
those from the pressure necessary to keep the source static).
Covariant conservation of the source on the brane allows one to
ascertain the source pressure, $p(r)$, given the source density
$\rho(r)$:
\be
	{p_g}' = - n'\rho_g\ .
\label{covariant}
\ee
One can show that the $G_{zr}$--component of the bulk is identically
zero on the brane when covariant conservation of the matter source and
the brane boundary conditions Eqs.~(\ref{branebc}) are satisfied.  This
observation is a restatement of one of the Bianchi identities.

From the brane boundary conditions Eqs.~(\ref{branebc2}), one may
eliminate $\dot{a}$ and $\dot{b}$, and arrive at an equation which
may be integrated imediately with respect to the $r$--coordinate.
The result yields
\be
	r^2 n' + ra + {1\over 2}r^3H^2 = R_g\ ,
\label{relation1}
\ee
where we have imposed a boundary condition requiring that the metric
not be singular at the origin.  Applying Eqs.~(\ref{relation1}) and
~(\ref{branebc2}) to the $G_{zz}$--component of
the bulk metric, one can again arrive at an equation that is dependent
only on variables on the brane itself which is integrable with respect
to the $r$--coordinate.
Defining the quantity
\be
	f(r) = rn' - {R'_g\over 2r} + {r^2H\over 2r_0}(2r_0H \pm 1)\ ,
\label{f}
\ee
the integral of the equations $G_{zz}=0$ on the brane yields
\be
	{4r_0^2\over r}f^2 + rf(3 \pm 2r_0H)
	- \left[{1\over 2}R_g
	  + r^3\left(2H^2 \pm {3H\over 2r_0}\right)\right] = 0\ ,
\label{relation2}
\ee
where we have applied a second spatial boundary condition requiring
that the metric not have spurious cusps at the origin.
Equation~(\ref{relation2}) has two solutions for $f(r)$.  We select
the solution that matches onto the proper background deSitter behavior
for large--$r$.  Then,
\be
	f(r) = {r^2\over 8r_0^2}\left[-(3\pm 2r_0H)
		+ 3(1\pm 2r_0H)\sqrt{1+{8r_0^2R_g(r)\over 9r^3}
		  {1\over (1\pm 2r_0H)^2}}~\right]\ .
\ee
Substituting this back into Eq.~(\ref{f}) and Eq.~(\ref{relation1}),
we may articulate expressions for $n(r)$ and $a(r)$.
\bea
	rn' &=& {R_g\over 2r}\left[1+ \delta(r)\right] - H^2r^2
\label{brane-n}     \\
	a &=& {R_g\over 2r}\left[1 - \delta(r)\right] + {1\over 2}H^2r^2\ ,
\label{brane-a}
\eea
where we have defined the quantity $\delta(r)$ such that
\be
	\delta(r) = {3r^3\over 4r_0^2R_g}(1\pm2r_0H)
		\left[\sqrt{1+{8r_0^2R_g\over 9r^3}{1\over (1\pm 2r_0H)^2}}
		- 1\right]\ .
\label{delta}
\ee
These expressions are valid on the brane when $r \ll r_0, H^{-1}$.  In both
expressions, the first term represent the usual Schwarzschild contribution
with a correction governed by $\delta(r)$ resulting from brane dynamics,
whereas the second term represents the leading cosmological contribution.
Let us try to understand the character of the corrections.

\section{Gravitational Regimes}

There are important asymptotic limits of physical relevance for the metric
on the brane Eqs.~(\ref{brane-n}--\ref{brane-a}).  We have two system
parameters, the crossover scale, $r_0$, and the cosmological horizon
radius, $H^{-1}$, and we wish to understand how spacetime geometry
differs when each is much larger than the other, as well as when they are
the same order of magnitude.

When $r_0H \ll 1$, we recover familiar results for a compact source in a
Minkowski background \cite{Gruzinov:2001hp,Porrati:2002cp}.  The
deSitter background becomes irrelevant at scales $r \ll r_0$.  A
characteristic distance $(r_0^2R_g)^{1/3}$ demarks an Einstein phase
close to the source from a Brans--Dicke ($\omega=0$) phase farther
away.

\subsection{Einstein Phase}

Cosmological effects become important to the metric
Eqs.~(\ref{brane-n}--\ref{brane-a}) when $r_0H \sim 1$ or $r_0H \gg 1$.
Typically, we are concerned with the details of the metric when the
influence of a given compact matter source dominates the local geometry.
The competition between the leading Schwarzschild term, $\sim R_g/r$,
versus the leading cosmological contribution, $\sim H^2r^2$, implies that
when
\be
	r \ll \left(R_g\over H^2\right)^{1/3}\ ,
\label{radius1}
\ee
the local source dominates the metric over the contributions from the
cosmological flow.  In this region, Eqs.~(\ref{brane-n}--\ref{brane-a})
reduce to
\bea
	n &=& -{R_g\over 2r} \pm \sqrt{R_g r\over 2r_0^2}
\label{Einstein-n}     \\
	a &=& {R_g\over 2r} \mp \sqrt{R_g r\over 8r_0^2}\ .
\label{Einstein-a}
\eea
Notice that, indeed, there is no explicit dependence on the parameter
governing cosmological expansion, $H$.  However, the sign of the
correction to the Schwarzschild solution is dependent on the global
properties of the cosmological phase.  The FLRW phase has a potential
steeper than Einstein gravity, whereas the self-accelerating phase has
a potential shallower than Einstein gravity.  Thus, we may ascertain
information about bulk, five-dimensional cosmological behavior from
testing details of the metric where naively one would not expect
cosmological contributions to be important.  Indeed, for example, one
can use local gravity tests to distinguish whether some local
brane vacuum energy or self-acceleration is the cause of today's
cosmic accelerated expansion.

\subsection{Weak Brane Phase}

Even when the cosmological flow dominates the metric, one can
still examine the perturbed effect a matter source has in this
region.  When, $r \gg (R_g H^{-2})^{1/3}$, while still in a region
well within the cosmological horizon ($r \ll H^{-1}$),
\bea
        n &=& -{R_g\over 2r}\left[1 + {1\over 3(1\pm 2r_0H)}\right]
                                  - {1\over 2}H^2r^2
\label{weakbrane-n}     \\
        a &=& {R_g\over 2r}\left[1 - {1\over 3(1\pm 2r_0H)}\right]
                                  + {1\over 2}H^2r^2\ .
\label{weakbrane-a}
\eea
This is the direct analog of the weak-brane phase one finds for
compact sources in Minkowski space.  The residual effect of the
matter source is a linearized scalar-tensor gravity with
Brans--Dicke parameter
\be
	\omega = \pm 3r_0H\ .
\label{BD}
\ee
Notice  that as  $r_0H \rightarrow  \infty$, we  recover  the Einstein
solution,  corroborating  results  found for  linearized  cosmological
perturbations  \cite{Deffayet:2002fn,Deffayet-talk}.   Moreover,
note that in the self-accelerating cosmological phase, the scalar
component couples repulsively,  though  recall  from  Eq.~(\ref{deS-rho})  that  $H  \ge r_0^{-1}$ in this phase.

\subsection{The Picture}

One can consolidate these results and show from
Eqs.~(\ref{brane-n}--\ref{delta}), that there exists a scale,
\be
	r_* = \left[{r_0^2R_g\over(1\pm 2r_0H)^2}\right]^{1/3}\ ,
\label{radius2}
\ee
inside of which the metric is dominated by Einstein but has
corrections which depend on the global cosmological phase, i.e.,
Eqs.~(\ref{Einstein-n}--\ref{Einstein-a}).  Outside this radius (but at
distances much smaller than both the crossover scale, $r_0$, and
the cosmological horizon, $H^{-1}$) the metric is weak-brane and
resembles a scalar-tensor gravity in the background of a deSitter
expansion, i.e., Eqs.~(\ref{weakbrane-n}--\ref{weakbrane-a}).

The picture one develops is that the metric is scalar-tensor, but with
a radially dependent coupling of the scalar to matter.  As one gets
closer to a source (as gravity gets stronger), the scalar coupling
gets suppressed.  There are two basic tests that we can pose.  First,
small corrections to the Newtonian potential using
Eq.~(\ref{Einstein-n}) and second, comparison of the ratio of the
gravitomagnetic to gravitoelectric (Newtonian) forces. This latter may
be encoded by the effective Brans--Dicke parameter, Eq.~(\ref{BD}).

\section{Physical Considerations}

\subsection{Constraints on $r_0$}

It is useful to review the constraints on the crossover scale, $r_0$.
The constraint may be written as a constraint on the fundamental
Planck scale: \be 10^{-3}~{\rm eV} \lesssim M \lesssim 1~{\rm GeV}\ .
\ee If $M$ is too small, quantum gravity effects become important at
short distances.  The lower bound is provided by constraints from
millimeter tests of Newtonian gravity and other constraints coming
from loss of energy into the extra dimension \cite{Dvali:2001gx}.
Cosmology provides the upper bound for $M$.  If $M$ is too large, the
crossover scale becomes too small to account for the relationship
between the observed Hubble scale and the independently measured
matter density \cite{Deffayet:2001pu,Deffayet:2002sp}.  These bounds
on the fundamental Planck scale correspond to
\be
	1~{\rm Gpc} \lesssim r_0 \lesssim 10^{34}~{\rm Gpc}\ .
\ee
An even more stringent case applies to the self-accelerating
cosmological phase.  In this phase, the Hubble scale is
bounded from below $H(t) > r_0^{-1}$, where at late times,
the universe is deSitter and $H \rightarrow r_0^{-1}$.  If one
goes further and postulates that the current cosmic acceleration
is caused entirely by this late-time self-acceleration, then,
using constraints from Type 1A supernovae \cite{Deffayet:2002sp},
the best fit for $r_0$ is
\be
	r_0 = 1.21_{-0.09}^{+0.09}H_0^{-1}\ ,
\ee
where $H_0$ is today's Hubble scale.  Taking
$H_0 \approx 70~{\rm km~s^{-1}Mpc^{-1}}$,
\be
        r_0 \approx 5~{\rm Gpc}\ .
\ee
We are particularly interested this possibility and, in this
Section, take $r_0$ to have this value.

\subsection{Gravitational Lensing}

The lensing of light by a compact matter source with metric
Eq.~(\ref{brane-n}--\ref{brane-a}) may be computed in the usual way.
The angle of deflection of a massless test particle is given by
\be
       \Delta\phi = \int dr~{J\over r^2}
		  {A\over \sqrt{{E^2\over N^2} - {J^2\over r^2}}}\ ,
\ee
where $E = N^2dt/d\lambda$ and $J = r^2d\phi/d\lambda$ are constants
of motion resulting from the isometries, and $d\lambda$ is the
differential affine parameter.  Then for any metric respecting the
condition Eq.~(\ref{relation1}), the angle of deflection is
\be
       \Delta\phi = \pi + 2b\int_b^{r_{\rm max}} 
       dr~{R_g(r) - H^2r^3/2\over r^2\sqrt{r^2-b^2}}\ ,
\ee
where $b$ is the impact parameter.  This result is equivalent to
Einstein, so we see that light deflection is unaffected by DGP
corrections.  This is consistent with the idea the DGP corrections
correspond to an anomalous spatially-dependent scalar coupling.
Since scalars do not couple to light, the trajectory of light in
a gravitational field should remain unaffected.  Then, lensing
measurements probe the true mass of a given matter distribution.
One can then compare that mass to the mass taken from assuming
the gravitational force is Newtonian.

\begin{figure}
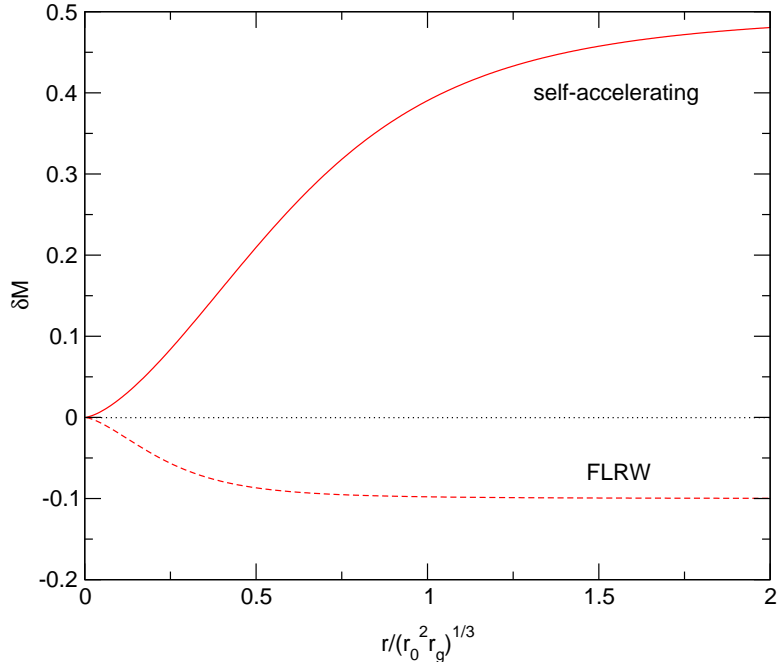
 \begin{center}\PSbox{mass.eps
hscale=50 vscale=50 hoffset=-60 voffset=-25}{3in}{3.1in}\end{center}
\caption{
Mass discrepancy, $\delta M$, for a static point source whose
Schwarzschild radius is $r_g$.  The solid curve is for a
self-accelerating background with $H = r_0^{-1}$.  The dashed curve is
for a FLRW background with $H = r_0^{-1}$.
}
\label{fig:mass}
\end{figure}

The mass discrepancy between the lensing mass (the actual mass) and that
determined from the Newtonian force may be read directly
from Eqs.~(\ref{brane-n}) and (\ref{delta}),
\be
       \delta M = {M_{\rm lens}\over M_{\rm Newt}}-1
       = {1\over 1+\delta(r)} - 1\ .
\ee
This ratio is depicted in Fig.~\ref{fig:mass} for both cosmological
phases.  When the mass is measured deep within the Einstein regime,
the mass discrepancy simplifies to
\be
       \delta M = \mp \left(r^3 \over 2r_0^2 R_g\right)^{1/3}\ .
\ee
Solar system measurements are too coarse to be able to resolve the DGP
discrepancy between lensing mass of the sun and its Newtonian mass.
The discrepancy $\delta M$ for the sun at ${\cal O}({\rm AU})$ scale
distances is approximately $10^{-11}$.  Limits on this discrepancy for
the solar system as characterized by the post-Newtonian parameter,
$\gamma-1$, are only constrained to be $< 3\times 10^{-4}$.

A possibly more promising regime may be found in galaxy clusters.  For
$10^{14} \rightarrow 10^{15}~M_\odot$ clusters, the scale
$(r_0^2R_g)^{1/3}$ has the range $6\rightarrow 14~{\rm Mpc}$.  For
masses measured at the cluster virial radii of roughly $1\rightarrow
3~{\rm Mpc}$, this implies mass discrepancies of and taking their
virial radii to be roughly $5\rightarrow 8\%$.  X-ray or
Sunyaev--Zeldovich (SZ) measurements are poised to map the Newtonian
potential of the galaxy clusters, whereas weak lensing measurements
can directly measure the cluster mass profile.  Unfortunately, these
measurements are far from achieving the desired precisions.  If one
can extend mass measurements to distances on the order of $r_0$,
Fig.~\ref{fig:mass} suggests discrepancies can be as large as $-10\%$
for the FLRW phase or even $50\%$ for the self-accelerating phase.

\subsection{Orbit Precession}

Imagine a body orbiting a mass source where $R_g(r) = r_g = {\rm
constant}$.  The perihelion precession per orbit may be determined
in the usual way
\be
       \Delta\phi = \int dr~{J\over r^2}
		  {AN\over \sqrt{E^2 - N^2\left(1+{J^2\over r^2}\right)}}\ ,
\ee
where $E = N^2dt/ds$ and $J = r^2d\phi/ds$ are again constants of
motion resulting from the isometries, and now $ds$ is the differential
proper time of the orbiting body.  Assuming a nearly circular orbit
and that we are deep within the Einstein regime (so that we may use
Eqs.~(\ref{Einstein-n}--\ref{Einstein-a})), then
\be
       \Delta\phi = 2\pi + {3\pi r_g\over r}
       \mp {3\pi\over 2}\left(r^3\over 2r_0^2r_g\right)^{1/2}\ .
\ee
The second term is the usual Einstein precession.  The last term is
the new anomalous precession due to DGP brane effects.  Note that
this correction is the same as one would get if one assumed a purely
Newtonian potential Eq.~(\ref{Einstein-n}) without spatial metric
effects.  The correction to the precession rate one expects from
DGP gravity is
\be
{d\over dt}{\Delta\phi}_{\rm DGP} = \mp {3\over 8r_0}
= \mp 5~\mu{\rm as/year}\ .
\label{corr-DGP}
\ee
Note that this result is independent of the source mass, implying that
this precession rate is a universal quantity dependent only on the
graviton's effective linewidth ($r_0^{-1}$) and the overall cosmological
phase.  Compare Eq.~(\ref{corr-DGP}) to the classic Einstein precession
correction for nearly circular orbits:
\be
{d\over dt}{\Delta\phi}_{\rm Einstein} =
    {3\over 2}\left(r_g^3\over 2r^5\right)^{1/2}\ .
\ee
For increasing $r$, the distance from the sun at which the DGP
correction begins to overtake the first Einstein correction is
$37~{\rm AU}$.

Nordtvedt \cite{Nordtvedt:ts} quotes precision for perihelion
precession at $430~\mu{\rm as/year}$ for Mercury and $10~\mu{\rm
as/year}$ for Mars.  Improvements in lunar ranging measurements
\cite{Williams:1995nq,lunar} suggest that the Moon will be
sensitive to the DGP correction Eq.~(\ref{corr-DGP}) in the near
future.  Also, BepiColombo, an ESA satellite being sent to Mercury at
the end of the decade, will also be sensitive to this correction
\cite{Milani:2002hw}.  Incidentally, it is interesting to contrast these
numbers with a precision of $4\times 10^4~\mu{\rm as/year}$ for the
rate of periastron advance in Binary Pulsar PSR~1913+16
\cite{Will:2001mx}.  The solar system seems to provide the most
promising means to constrain this anomalous precession from DGP
gravity.

\section{Concluding Remarks}

The braneworld theory of Dvali--Gabadadze--Porrati (DGP) is an
intriguing extension of Einstein gravity that exploits the possible
existence of infinite-volume, extra dimensions.  It is a theory where
the four-dimensional graviton is effectively metastable, and 
provides a novel alternative to conventional explanations of the
dark energy that is responsible for today's cosmic acceleration.

In this paper, we detailed the solution of static, spherical matter
sources in the background of deSitter cosmology for DGP gravity.
The gravitational field of a matter source exhibits important
dependences on cosmology.  Residual dependences on the full
five-dimensional cosmological phase also exist in the regime
deep in the gravity well of the matter source where the effects of
cosmology are ostensibly irrelevant.  These residual dependences
allow one to use local (e.g., solar system) measurements of the
gravitational field to ascertain details of the global cosmology.

In DGP gravity, we find that massless test particles can probe the
true mass of a matter source, whereas tests of the source's
Newtonian force leads to discrepancies with general relativity.
These discrepancies translate into a universal anomalous
precession, as large as $\pm 5~\mu{\rm as/year}$, suffered by all
orbiting bodies.  The numerical value of this anomalous precession
is dependent only on the graviton's effective linewidth and the
global geometry of the five-dimensional cosmology.  Current
constraints on Mars' orbit are on the threshold of being sensitive to
this anomaly \cite{Nordtvedt:ts}.  Future improvements in lunar
ranging \cite{lunar} as well as data from satellite missions at the
end of the decade \cite{Milani:2002hw} should be sensitive to
possible corrections due to DGP and gravitational leakage into
extra dimensions.

\acknowledgements

The authors would like to thank A.~Gruzinov for crucial insights into
the Minkowski background case, particularly those not found in
Ref.~\cite{Gruzinov:2001hp}.  We are also grateful to L. Krauss for
key suggestions and conversations.  The authors would also like to
thank A.~Babul, C.~Deffayet, G.~Dvali, P.~Gondolo, and G.~Kofinas for
helpful discussions.  This work is sponsored by DOE Grant
DEFG0295ER40898 and the CWRU Office of the Provost.

\appendix
\section*{}

In order to see why Eq.~(\ref{assumption}) is a reasonable approximation,
we need to explore the full solution to the bulk Einstein equations,
\be
	G_{AB}(r,z) = 0\ ,
\ee
satisfying the brane boundary conditions, Eqs.~(\ref{branebc}), as well as
specifying that the metric approach the deSitter solution
Eqs.~(\ref{deS-N}--\ref{deS-B}) for large values of $r$ and $z$, i.e., far
away from the compact matter source.

First, it is convenient to consider not only the components of the Einstein
tensor Eqs.~(\ref{5dEinstein}), but also the following components of the
bulk Ricci tensor (which also vanishes in the bulk):
\bea
R_t^t &=& {1\over A^2}\left[{N''\over N} - {N'\over N}{A'\over A}
  + 2{N'\over N}{B'\over B}\right]
+ \left[{\ddot{N}\over N} + {\dot{N}\over N}{\dot{A}\over A}
  + 2{\dot{N}\over N}{\dot{B}\over B}\right]
\label{5dRicci-tt}     \\
R_z^z &=& {\ddot{N}\over N} + {\ddot{A}\over A} + {2\ddot{B}\over B}\ .
\label{5dRicci-zz}
\eea
We wish to take $G_{zr}=0$, $G_z^z=0$, and $R_z^z=0$ and derive
expressions for $A(r,z)$ and $B(r,z)$ in terms of $N(r,z)$.  Only two of
these three equations are independent, but it is useful to use all
three to ascertain the desired expressions.

Since we are only interested in metric when $r,z \ll r_0, H^{-1}$ for
a weak matter source, we may rewrite the necessary field equations
using the expressions Eqs.~(\ref{linearize}).  Since the functions,
$\{n(r,z),a(r,z),b(r,z)\}$ are small, we need only keep nonlinear
terms that include $z$--derivatives.  The brane boundary conditions,
Eqs.~(\ref{branebc}), suggest that $\dot{a}$ and $\dot{b}$ terms may
be sufficiently large to warrant inclusion of their subleading
contributions.  It is these $z$--derivative nonlinear terms
that are crucial to the recover of Einstein gravity near the matter
source.  If one neglected these bilinear terms as well, one would
revert to the linearized, weak-brane solution
(cf. Ref.~\cite{Gruzinov:2001hp}).

Integrating Eq.~(\ref{5dRicci-zz}) twice with respect to the
$z$--coordinate, we get
\be
	n + a + 2b = zg_1(r) + g_2(r)\ ,
\label{app-relation1}
\ee
where $g_1(r)$ and $g_2(r)$ are to be specified by the brane
boundary conditions, Eqs.~(\ref{branebc}), and the residual
gauge freedom $\delta b(r)|_{z=0} = 0$, respectively.
Integrating the $G_{zr}$--component of the bulk Einstein tensor
Eqs.~(\ref{5dEinstein}) with respect to the $z$--coordinate yields
\be
	r\left(n + 2b\right)' - 2\left(a-b\right) = g_3(r)\ .
\label{app-relation2}
\ee
The functions $g_1(r)$, $g_2(r)$, and $g_3(r)$ are not all
independent, and one can ascertain their relationship with one
another by substituting Eqs.~(\ref{app-relation1}) and
(\ref{app-relation2}) into the $G_z^z$ bulk equation.  If one can
approximate $\dot{n} = \mp H$ for all $z$, then one can see
that  $G_{zr}=0$, $G_z^z=0$, and $R_z^z=0$ are all consistently
satisfied by Eqs.~(\ref{app-relation1}) and (\ref{app-relation2}),
where the functions $g_1(r)$, $g_2(r)$, and $g_3(r)$ are
determined at the brane using Eqs.~(\ref{brane-n}) and
(\ref{brane-a}) and the residual gauge freedom
$b(r)|_{z=0} = 0$:
\bea
	g_1(r) &=& \mp 4H - {r_0\over r^2}\left(R_g\delta\right)'
\label{g1}	\\
	g_2(r) &=& {R_g\over 2r}(1-\delta)
		+ \int_0^r dr~{R_g\over r^2}(1+\delta)
\label{g2}	\\
	g_3(r) &=& {R_g\over 2r}(1-3\delta) - 2H^2r^2\ ,
\label{g3}
\eea
where we have used the function $\delta(r)$, defined in
Eq.~(\ref{delta}).  Using Eqs.~(\ref{app-relation1}--\ref{g3}), we
now have expressions for $a(r,z)$ and $b(r,z)$ completely in
terms of $n(r,z)$ for all $(r,z)$.

Now we wish to find $n(r,z)$ and to confirm that
$\dot{n} = \mp H$ is a good approximation everywhere of interest.
Equation~(\ref{5dRicci-tt}) becomes
\be
	n'' + {2n'\over r} + \ddot{n} = \pm H[g_1(r)\pm H]\ ,
\ee
where again we have neglected contributions if we are only
concerned with $r,z \ll r_0, H^{-1}$.  Using the expression
Eq.~(\ref{g1}), we find
\be
	n'' + {2n'\over r} + \ddot{n} = -3H^2
		\mp {r_0H\over r^2}\left[R_g\delta(r)\right]'\ .
\ee
Then, if we let
\be
	n = 1 \mp Hz - {1\over 2}H^2r^2
			\mp r_0H\int_0^r dr~{1\over r^2}R_g(r)\delta(r)
			+ \delta n(r,z)\ ,
\label{app-relation3}
\ee
where $\delta n(r,z)$ satisfies the equation
\be
	\delta n'' + {2\delta n'\over r} + \ddot{\delta n} = 0\ ,
\label{potential}
\ee
we can solve Eq.~(\ref{potential}) by requiring that $\delta n$
vanish as $r,z\rightarrow \infty$ and applying the condition
\be
	r~\delta n'|_{z=0} = {R_g\over 2r}
	\left[1 + (1\pm 2r_0H)\delta(r)\right]\ ,
\label{brane-dn}
\ee
on the brane as an alternative to the appropriate brane boundary
condition for $\delta n(r,z)$ coming from a linear combination of
Eqs.~(\ref{branebc}).  We can write the solution explicitly:
\be
        \delta n(r,z) = \int_0^{\infty}dk~c(k)e^{-kz}\sin kr\ ,
\ee
where
\be
        c(k) = {2\over \pi}\int_0^\infty dr~r\sin kr
	        \left.\delta n\right|_{z=0}(r)\ .
\ee
We can then compute $\dot{\delta n}|_{z=0}$, arriving at the bound
\be
	\dot{\delta n}|_{z=0} \lesssim
		{1\over r}\int_0^r dr~{R_g(r)\over r^2}\ ,
\ee
for all $r \ll r_0, H^{-1}$.  Then,
\be
	\dot{n}|_{z=0} = \mp H + \dot{\delta n}|_{z=0}\ .
\label{dotn}
\ee
When the first term in Eq.~(\ref{dotn}) is much larger than the
second, Eq.~(\ref{assumption}) is a good approximation.  When
the two terms in Eq.~(\ref{dotn}) are comparable or when the
second term is much larger than the first, neither term is
important in the determination of Eqs.~(\ref{brane-n}) and
(\ref{brane-a}).  Thus, Eq.~(\ref{assumption}) is still a safe
approximation.

One can confirm that all the components of the five-dimensional
Einstein tensor, Eqs.~(\ref{5dEinstein}), vanish in the bulk using
field variables satisfying the relationships
Eqs.~(\ref{app-relation1}), (\ref{app-relation2}), and
(\ref{app-relation3}).  The field variables $a(r,z)$ and $b(r,z)$ both
have terms that grow with $z$, stemming from the presence of the
matter source.  However, one can see that with the following
redefinition of coordinates:
\bea
	R &=& r - zr_0{R_g\delta\over r^2}	\\
	Z &=& z + \int_0^r dr~ {R_g\delta\over r^2}\ ,
\eea
that to leading order as $z \rightarrow H^{-1}$, the
desired $Z$--dependence is recovered for $a(R,Z)$ and $b(R,Z)$
(i.e., $\mp HZ$), and the Newtonian potential takes the form
\be
	n(R,Z) = \mp HZ -  {1\over 2}H^2R^2 + \cdots\ .
\ee
Thus, we recover the desired asymptotic form for the metric
of a static, compact matter source in the background of a
deSitter expansion.


\begin{thebibliography}{99}


\bibitem{Dvali:2000hr}
G.~Dvali, G.~Gabadadze and M.~Porrati,
Phys.\ Lett.\ B {\bf 485}, 208 (2000).

\bibitem{Dvali:2001gm}
G.~R.~Dvali, G.~Gabadadze, M.~Kolanovic and F.~Nitti,
Phys.\ Rev.\ D {\bf 64}, 084004 (2001).

\bibitem{Dvali:2001gx}
G.~R.~Dvali, G.~Gabadadze, M.~Kolanovic and F.~Nitti,
Phys.\ Rev.\ D {\bf 65}, 024031 (2002).

\bibitem{Deffayet}
C.~Deffayet,
Phys.\ Lett.\ B {\bf 502}, 199 (2001).

\bibitem{Deffayet:2001pu}
C.~Deffayet, G.~R.~Dvali and G.~Gabadadze,
Phys.\ Rev.\ D {\bf 65}, 044023 (2002).

\bibitem{Deffayet:2002sp}
C.~Deffayet, S.~J.~Landau, J.~Raux, M.~Zaldarriaga and P.~Astier,
Phys.\ Rev.\ D {\bf 66}, 024019 (2002).

\bibitem{Alcaniz:2002qh}
J.~S.~Alcaniz,
Phys.\ Rev.\ D {\bf 65}, 123514 (2002).

\bibitem{Jain:2002di}
D.~Jain, A.~Dev and J.~S.~Alcaniz,
Phys.\ Rev.\ D {\bf 66}, 083511 (2002).

\bibitem{Alcaniz:2002qm}
J.~S.~Alcaniz, D.~Jain and A.~Dev,
Phys.\ Rev.\ D {\bf 66}, 067301 (2002).

\bibitem{Lue:2002fe}
A.~Lue,
arXiv:hep-th/0208169.

\bibitem{Deffayet:2001uk}
C.~Deffayet, G.~R.~Dvali, G.~Gabadadze and A.~I.~Vainshtein,
Phys.\ Rev.\ D {\bf 65}, 044026 (2002).

\bibitem{Lue:2001gc}
A.~Lue,
Phys.\ Rev.\ D {\bf 66}, 043509 (2002).

\bibitem{Gruzinov:2001hp}
A.~Gruzinov,
arXiv:astro-ph/0112246.

\bibitem{Porrati:2002cp}
M.~Porrati,
Phys.\ Lett.\ B {\bf 534}, 209 (2002).

\bibitem{Middleton:2002qa}
C.~Middleton and G.~Siopsis,
arXiv:hep-th/0210033.

\bibitem{Kofinas:2001qd}
G.~Kofinas, E.~Papantonopoulos and I.~Pappa,
arXiv:hep-th/0112019.

\bibitem{Kofinas:2002gq}
G.~Kofinas, E.~Papantonopoulos and V.~Zamarias,
arXiv:hep-th/0208207.

\bibitem{Deffayet:2002fn}
C.~Deffayet,
arXiv:hep-th/0205084.

\bibitem{Deffayet-talk}
C. Deffayet, talk given at ``Peyresq Physics 7''
Conference, June 2002, Peyresq (France).

\bibitem{Nordtvedt:ts}
K.~Nordtvedt,
Phys.\ Rev.\ D {\bf 61}, 122001 (2000).

\bibitem{Williams:1995nq}
J.~G.~Williams, X.~X.~Newhall and J.~O.~Dickey,
Phys.\ Rev.\ D {\bf 53}, 6730 (1996).

\bibitem{lunar}
G.~Dvali, A.~Gruzinov, and M.~Zaldarriaga, hep-ph/0212069.

\bibitem{Milani:2002hw}
A.~Milani, D.~Vokrouhlicky, D.~Villani, C.~Bonanno and A.~Rossi,
Phys.\ Rev.\ D {\bf 66}, 082001 (2002).

\bibitem{Will:2001mx}
C.~M.~Will,
Living Rev.\ Rel.\  {\bf 4}, 4 (2001)
[arXiv:gr-qc/0103036].

\end{thebibliography}
\end{document}